\documentclass[12pt]{article}

\usepackage{latexsym,amsmath,epsfig}

\DeclareFontFamily{OT1}{rsfs10}{}
\DeclareFontShape{OT1}{rsfs10}{m}{n}{ <-> rsfs10 }{}
\DeclareMathAlphabet{\mathscript}{OT1}{rsfs10}{m}{n}

\newcommand{\mbf}[1]{\mathbf{#1}}

\newcommand{\ns}{\normalsize}

\newcommand{\Ds}{\not\!\!D}

\newcommand{\ZZ}{{\bf Z}}

\newcommand{\PP}{{\bf CP}}

\def\l{\lambda}

\def\s{\sigma}

\def\cC{{\cal C}}

\def\cL{{\cal L}}

\def\cM{{\cal M}}
\def\cN{{\cal N}}

\begin{document}

%%%%%%%%%%%%%%%%%%%%%%%%%%%%%%%%%%%%%%%%%%%%%%%%%%%%%%%%%%%%%%%%%%%%%%

\begin{titlepage}

\vspace{-5cm}

\title{
   \hfill{\ns UPR-823T, OUTP-98-80P, PUPT-1822} \\[3em]
   {\LARGE Non-Perturbative Vacua and Particle Physics in M-Theory}
       \\[1em] } 
\author{
   {\ns Ron Donagi$^1$, 
      Andr\'e Lukas$^2$, 
      Burt A.~Ovrut$^3$\setcounter{footnote}{0}\thanks{
          Supported in part by a Senior Alexander 
          von Humboldt Award.} \ and Daniel Waldram$^4$} \\[0.5em]
   {\it\ns $^1$Department of Mathematics, 
      University of Pennsylvania} \\[-0.5em]
   {\it\ns Philadelphia, PA 19104--6395, USA}\\ 
   {\it\ns $^2$Department of Physics, Theoretical Physics, 
       University of Oxford} \\[-0.5em]
   {\it\ns 1 Keble Road, Oxford OX1 3NP, United Kingdom}\\ 
   {\it\ns $^3$Department of Physics, University of Pennsylvania} \\[-0.5em]
   {\it\ns Philadelphia, PA 19104--6396, USA}\\ 
   {\it\ns $^4$Department of Physics, Joseph Henry Laboratories,}\\[-0.5em]
   {\it\ns Princeton University, Princeton, NJ 08544, USA}}
\date{}

\maketitle

\begin{abstract}
In this letter, we introduce a general theory for the construction of
particle physics theories, with three families and realistic gauge
groups, within the context of heterotic M-theory. This is achieved
using semi-stable holomorphic gauge bundles over elliptically fibered
Calabi--Yau three-folds. Construction of realistic theories is
facilitated by the appearance of non-perturbative five-branes in the
vacuum. The complete moduli space of these five-branes is computed and
their worldvolume gauge theory discussed. It is shown, within the
context of holomorphic gauge bundles, how grand unified gauge groups
can be spontaneously broken to the gauge group of the standard
model. These ideas are illustrated in an explicit $SU(5)$ three-family
example. 

\end{abstract}

\thispagestyle{empty}

\end{titlepage}

%%%%%%%%%%%%%%%%%%%%%%%%%%%%%%%%%%%%%%%%%%%%%%%%%%%%%%%%%%%%%%%%%%%%%%%

In seminal work, Horava and Witten~\cite{HW1} and Witten~\cite{HW2}
showed that, by compactifying $M$-theory on an orbifold interval,
$S^{1}/Z_{2}$, times a Calabi--Yau three-fold, $X$, ${\cal{N}}=1$
supersymmetric $E_{8} \times E_{8}$ gauge theories with chiral
fermions can arise in four dimensions. In~\cite{losw1,losw2}, the
effective five-dimensional heterotic $M$-theory was constructed and its static
vacuum shown to be an exact pair of BPS three-branes located at the
orbifold planes. More recently~\cite{nse}, this work was extended by
the inclusion in the vacuum of five-branes wrapped on holomorphic
curves in the Calabi--Yau three-fold and located at arbitrary points
on the orbifold interval. It was emphasized that, in the context of
$M$-theory, the standard embedding of the spin connection into the
gauge connection plays no special role. Naturally, one should consider
arbitrary non-standard embeddings. These vacua with five-branes and
non-standard embeddings were called non-perturbative heterotic
$M$-theory vacua. 

In this letter, we give concrete form to these ideas by showing how to
explicitly construct non-perturbative vacua. 
A more complete description will be given in~\cite{future}.
First, the difficult problem of non-standard embedding is solved using
techniques introduced in~\cite{FMW,Ron} for the construction of
holomorphic vector bundles over elliptically fibered Calabi--Yau
three-folds. Second, the requirement that the vacua be anomaly free
and $\cN=1$ supersymmetric leads to a cohomological condition that
fixes the homology class of the five-branes in terms of Chern classes
of the gauge bundles and the Calabi--Yau tangent space. In this paper,
we allow for the presence of five-branes in the vacua. The inclusion
of five-branes greatly facilitates the solution of the cohomology
condition and, hence, allows us to find many suitable vacua. We note 
that this construction does not require going to the M-theory
limit we will consider here. It is equally applicable as a new way of
building vacua with non-standard embeddings in the weakly coupled
limit, though any five-brane states would have no perturbative string
description.

Using our approach, we can explicitly construct non-perturbative vacua
describing realistic $\cN=1$ particle physics theories in four
dimensions. The phenomenological requirements we will use are (i)
that the low-energy gauge group be a suitable grand unified group,
such as $SU(5)$, $SO(10)$ or $E_{6}$, and (ii) that there be three
families. This leads us to choose the background gauge fields to lie
in an $SU(n)$ subgroup of $E_8$ and puts a constraint on the third
Chern class of the gauge bundles. The form of this constraint for
elliptically fibered Calabi--Yau three-folds was first presented
in~\cite{Andreas,Curio} and some of its solutions were given
in~\cite{Curio}. One then chooses a particular Calabi--Yau three-fold
and finds the relevant $SU(n)$ gauge bundle which is a solution of the
three-family condition. That this be a true $SU(n)$ bundle provides a
second constraint. The class of the five-branes is then calculated
from the cohomology condition. While the mathematics may seem
involved, this procedure can be reduced to a set of rules involving
properties of the base manifold of the Calabi--Yau three-fold, and a
single line bundle and discrete parameter describing the bundle. 

From the low-energy point of view, these theories have, in their
simplest form, a visible sector with a gauge group such as $SU(5)$,
$SO(10)$ or $E_6$ and a hidden sector with an unbroken $E_8$ gauge
group. In addition, the five-branes act as a further set of hidden
sectors, generically with unitary gauge groups on their worldvolumes. 
In five dimensions, each such hidden sector lives on a separate domain
wall at a different point in the orbifold interval. 
We find that many realistic particle physics vacua can be constructed this
way. In this paper, we consider one example, with gauge group $SU(5)$, in
detail. In particular, we give a complete description of the
five-branes in this example, calculating their moduli
space, the genus of the five-brane curve within the Calabi--Yau
manifold and discussing the form of the low-energy theory on each
five-brane worldvolume. 

Having constructed grand unified theories, it is essential to show how
these theories are spontaneously broken to the standard model. In this
paper, we will demonstrate that this is indeed possible, within the
context of holomorphic bundles over elliptically fibered Calabi--Yau
three-folds, using Wilson lines. This requires constructing
Calabi--Yau spaces with non-trivial fundamental group. 

Finally, we note that the authors of~\cite{FMW} were primarily
concerned with  understanding the duality between heterotic theory and
F-theory compactifications. In this letter, we will not discuss this
duality. Here, we simply remark that the non-perturbative vacua
discussed in this letter are more general than those usually
considered, since the orientation of the five-branes is such that they
do not, in general, map under duality to three-branes in F-theory. 
 
\bigskip

We will consider vacuum states of $M$-theory with the following structure.
The space-time structure of the vacuum is chosen to be that of 
Horava and Witten~\cite{HW1,HW2} and
Witten~\cite{W}, to lowest
order in the expansion parameter $\kappa^{2/3}$.
\begin{itemize}
\item Space-time is taken to have the form
\begin{equation}
   M_{11}=M_{4} \times S^{1}/Z_{2} \times X
\label{eq:1}
\end{equation}
where $M_{4}$ is four-dimensional Minkowski space, $S^1/Z_2$ is a
one-dimensional orbifold and $X$ is a smooth Calabi--Yau
three-fold. 
\end{itemize}
The vacuum space-time structure becomes more complicated
at the next order in $\kappa^{2/3}$, but this ``deformation''
can be viewed as arising as the static vacuum of the five-dimensional
effective theory~\cite{losw1,losw2} and, hence, need not concern us
here.

The $Z_2$ orbifold projection necessitates the introduction, on each
of the two orbifold planes, of an $\cN=1$, $E_8$ Yang-Mills
supermultiplet which is required for anomaly cancelation. 
The gauge field structure of these vacua is encoded in the gauge
bundle. To be compatible with four
preserved supercharges in four dimensions, the
gauge bundle on each orbifold plane must
be a semi-stable, holomorphic bundle with the structure group being the
complexification $E_{8\mathbf{C}}$ of $E_8$.
We will denote any group $G$ and its
complexification $G_{\mathbf{C}}$ simply as $G$, letting context dictate 
which group is being referred to. These semi-stable,
holomorphic gauge bundles are, a priori, allowed to be arbitrary in
all other respects. In particular, there is no requirement that the
spin-connection of the Calabi--Yau three-fold be embedded into an
$SU(3)$ subgroup of the gauge connection of one of the $E_8$ bundles,
the so-called standard embedding. This generalization to arbitrary
semi-stable holomorphic gauge bundles is what is referred to as
non-standard embedding. These terms are somewhat irrelevant in the 
context of M-theory, where no choice of embedding can ever set the
entire eleven-dimensional four-form field strength to zero. For this
reason, we will simply refer to arbitrary semi-stable holomorphic
$E_8$ gauge bundles. It is clear that we can restrict the transition
functions to be elements of any subgroup $G$ of $E_8$, such as
$G=SU(n)$ or $Sp(n)$. Hence, we choose the following gauge structure
for the vacua.  
\begin{itemize} 
\item There is a semi-stable holomorphic gauge bundle $V_{i}$ with fiber
group $G_{i} \subseteq E_{8}$ over the Calabi-Yau three-fold on the
$i$-th orbifold fixed plane for $i=1,2$. The structure groups $G_{1}$ and
$G_{2}$ of the two bundles can be any subgroups of $E_{8}$ and need
not be the same.
\end{itemize}

In addition, as discussed in~\cite{W,nse}, we can allow for the
presence of five-branes located at points throughout the orbifold
interval. The five-branes will preserve $\cN=1$ supersymmetry,
provided they are wrapped on holomorphic two-cycles within $X$ and
otherwise span the flat Minkowski space $M_4$~\cite{W,bbs,vb}. The
inclusion of five-branes is essential for a complete discussion of
M-theory vacua. The reason for this is that, given a Calabi--Yau
three-fold background, the presence of five-branes allows one to
construct large numbers of gauge bundles that would otherwise be
disallowed~\cite{nse}. 
\begin{itemize}
\item We allow for the presence of five-branes in the vacuum, which
are wrapped on holomorphic curves $W$ within $X$.
\end{itemize}

The requirements of gauge and gravitational anomaly cancelation on
the two orbifold fixed planes, as well as anomaly cancelation on each
five-brane worldvolume, necessitates the addition of four-form sources
to the four-form field strength Bianchi identity. Integrate this
Bianchi identity over any five-cycle which spans the orbifold
interval together with an arbitrary four-cycle ${\cal{C}}_{4}$ in the
Calabi-Yau three-fold. Since $dG$ is exact and the cycle is compact,
this integral must vanish. This means the sources are subject to the
following condition. 
\begin{itemize}
\item The Calabi-Yau three-fold, the gauge bundles and the five-branes
are subject to the cohomological constraint
\begin{equation}
   c_{2}(V_{1}) + c_{2}(V_{2}) + [W] = c_{2}(TX)
\label{eq:2}
\end{equation}
where $c_{2}(V_{i})$ and
$c_{2}(TX)$ are the second Chern classes of the gauge bundle $V_{i}$
and the tangent bundle $TX$ respectively and $[W]$ is the cohomology
class associated with the five-brane curves $W$.
\end{itemize}
Note that integrating this constraint over an arbitrary four-cycle
${\cal{C}}_{4}$ yields the expression 
\begin{equation}
   n_{1}({\cal{C}}_{4})+n_{2}({\cal{C}}_{4})+n_{5}({\cal{C}}_{4})=
       n_{R}({\cal{C}}_{4})
\label{eq:3}
\end{equation}
which states that the net magnetic charge on the four-cycle $\cC_4$
must vanish, so that the sum of the number of gauge instantons on the
two orbifold planes, plus the sum of the five-brane magnetic charges,
must equal the instanton number for the Calabi-Yau tangent bundle, a
number which is fixed once the Calabi-Yau three-fold is chosen. Vacua
of this type will be referred to as non-perturbative heterotic
M-theory vacua. 

\bigskip

The discussion given thus far is completely generic, in that it
applies to any Calabi-Yau three-fold and any gauge bundles that can be
constructed over it. However, realistic particle physics theories
require the explicit construction of these gauge bundles. Here, we
will present a formalism for choosing appropriate semi-stable
holomorphic gauge bundles with structure groups $G_{1}$ and $G_{2}$ over
the two orbifold fixed planes.
In this letter, for specificity, we will restrict the structure groups to
be 
\begin{equation}
   G_{i}= SU(n_{i})
\label{eq:18}
\end{equation}
for $i=1,2$. Other structure groups, such as $Sp(n)$ or exceptional
groups, will be discussed elsewhere. 

Our explicit bundle constructions~\cite{FMW,Ron} will be achieved over
the restricted, but rich, set of elliptically fibered Calabi-Yau
three-folds. These three-folds are known to be the simplest Calabi-Yau
spaces on which one can explicitly construct bundles, compute Chern
classes, moduli spaces and so on. This makes them a compelling choice
for the construction of concrete particle physics theories. Having
constructed the bundles, one can explicitly calculate the gauge bundle
Chern classes $c_{2}(V_{i})$ for $i=1,2$, as well as the tangent
bundle Chern class $c_{2}(TX)$. Having done so, one can then find the
class $[W]$ of the five-branes using the cohomology
condition~\eqref{eq:2}.

Elliptically fibered Calabi--Yau vacua have the following properties.
\begin{itemize}
\item An elliptically fibered Calabi--Yau three-fold $X$ is composed
of a two-fold base $B$, elliptic curves (that is tori) $E_{b}$ fibered
over each point $b \in B$ and an analytic map $\pi:X \rightarrow
B$. In this letter, we will assume there is a global section $\sigma$. 
\end{itemize}
The elliptic fibration is characterized by a single line bundle $\cL$
over $B$. The condition that the first Chern class of the tangent
bundle $TX$ of the Calabi--Yau three-fold $X$ vanish implies that
$\cL=-K_{B}$, where $K_{B}$ is the canonical bundle of the base
$B$. The fact that $\cL$ cannot be arbitrary limits the number of
possible bases~\cite{Grassi,MV}.
\begin{itemize}
\item If the base is smooth and preserves only $\cN=1$ supersymmetry in
four-dimensions, then $B$ is restricted to the following manifolds: 
an Enriques surface, a del Pezzo surface, a Hirzebruch surface, or a
blow up of a Hirzebruch surface. 
\item  The second Chern class of the holomorphic tangent bundle of $X$
is given by~\cite{FMW}
\begin{equation}
   c_2(TX) = c_2(B) + 11 c_1(B)^2 + 12 \s c_1(B)
\label{eq:add16}
\end{equation}
where $c_1(B)$ and $c_2(B)$ are the first and second Chern classes
of the tangent bundle of the base $B$.
\end{itemize}

Each gauge bundle over the elliptically fibered Calabi--Yau three-fold
has the following properties.
\begin{itemize}
\item  To specify a semi-stable $SU(n)$ gauge bundle $V$, we need to
fix a spectral cover and a line bundle $\cN$ over it. The class of the
spectral cover is itself specified in terms of a second line bundle
$\cM$ on the base $B$. The relevant quantities associated with $\cM$
and $\cN$ are their first Chern classes
\begin{equation}
   \eta= c_{1}(\cM)
\label{eq:add17}
\end{equation}
and $c_{1}(\cN)$ respectively. The class $c_{1}(\cN)$ can be
expressed in terms of $n, \sigma, \eta$ and $c_{1}(B)$, and an
additional rational coefficient $\lambda$.
\item  The condition that $c_{1}(\cN)$ be an integer leads to the
constraints on $\eta$ and $\lambda$ given by
\begin{align}
   & \text{$n$ is odd},  \quad \l = m+\frac{1}{2} 
\label{eq:add18} \\
   & \text{$n$ is even},  \quad 
      \l = m, \quad \eta = c_1(B) \!\!\mod 2 
\label{eq:add19}
\end{align}
where $m$ is an integer.
\item  The relevant Chern classes of an $SU(n)$ gauge bundle $V$ are
given by~\cite{FMW,Curio}
\begin{align}
   c_1(V) &= 0 \label{c1} \\
   c_2(V) &= \eta \s - \frac{1}{24} c_1(B)^2 \left(n^3 - n\right) 
              + \frac{1}{2} \left(\l^2 - \frac{1}{4}\right) n \eta 
                      \left(\eta - nc_1(B)\right) \label{c2} \\
   c_3(V) &= 2 \l \s \eta \left( \eta - nc_1(B) \right) \label{c3}
\end{align}
\end{itemize}

\bigskip

What gauge bundles should we choose in order to construct realistic
particle physics models? The simplest case is to choose an arbitrary
semi-stable $SU(n)$ gauge bundle $V_{1}$, which we henceforth call
$V$, on the first orbifold plane, but always take the gauge bundle
$V_{2}$ to be trivial. Physically, this corresponds to allowing
observable sector gauge groups to be subgroups, such as $SU(5)$,
$SO(10)$ or $E_{6}$, of $E_{8}$ but leaving the hidden sector $E_{8}$
gauge group unbroken.  These gauge groups arise as the commutants of
the $SU(n)$ structure groups. Explicitly, we have the following maximal
subgroups of $E_8$ together with the decompositions of the adjoint
representation of $E_{8}$ under those subgroups. 
\begin{equation}
\begin{aligned}
   SU(3) \times E_6 : & \quad
      \mbf{248} = (\mbf{8},\mbf{1}) \oplus (\mbf{1},\mbf{78}) \oplus
           (\mbf{3},\mbf{27}) \oplus (\mbf{\bar{3}},\mbf{\bar{27}}) \\
   SU(4) \times SO(10) : & \quad
      \mbf{248} = (\mbf{15},\mbf{1}) \oplus (\mbf{1},\mbf{45}) \oplus
           (\mbf{4},\mbf{16}) \oplus (\mbf{\bar{4}},\mbf{\bar{16}}) \\
   SU(5) \times SU(5) : & \quad
      \mbf{248} = (\mbf{24},\mbf{1}) \oplus (\mbf{1},\mbf{24}) \oplus
           (\mbf{10},\mbf{5}) \oplus (\mbf{\bar{10}},\mbf{\bar{5}}) \oplus
           (\mbf{5},\mbf{\bar{10}}) \oplus (\mbf{\bar{5}},\mbf{10})
\end{aligned}
\label{subgroup}
\end{equation}
Thus, for example, the observable sector gauge group $SO(10)$ arises as the
subgroup of $E_{8}$ commutant with the structure group $SU(4)$.
Note that choosing $V_2$ to be trivial is done only for
simplicity. Our formalism also allows a complete analysis of the
general case where the hidden sector $E_{8}$ gauge group is broken by
a non-trivial bundle $V_{2}$. 

With this choice of gauge bundle, it follows from  equation~\eqref{eq:2} that
the cohomology class associated with the five-branes is given by
\begin{equation}
   [W]=c_{2}(TX)-c_{2}(V)
\label{eq:x5}
\end{equation}
Inserting the expressions ~\eqref{eq:add16} and ~\eqref{c2} into this
equation allows one to explicitly determine the five-brane class $[W]$. Since
we are interested in the structure of the five-brane curves, we find
it expedient to use Poincar\'e duality to rewrite cohomological
expressions in terms of homology classes. In this context, we must
introduce the homology class of the fiber, which is denoted by $F$. If
our Calabi--Yau threefold is sufficiently general, we find the
following. 
\begin{itemize}
\item  The homology class associated with the five-branes is specifically of
the form 
\begin{equation}
 [W]= W_{B} + a_{f}F
 \label{eq:x16}
\end{equation}
where $W_B$ is the homology class in the base, 
\begin{gather}
   W_{B} = 12c_{1}(B)-\eta 
\label{eq:x17} \\
   a_{f} = c_{2}(B) 
             + c_{1}(B)^{2}\left(11+\frac{1}{24}(n^{3}-n)\right) 
             - \frac{n}{2}\left(\lambda^{2}-\frac{1}{4}\right)
                  \eta \left( \eta - nc_1(B) \right)
\label{eq:x18}
\end{gather}
and $c_{1}(B)$ and $c_{2}(B)$ are the first and second Chern
classes of the base $B$. 
\end{itemize}

Each fivebrane must be wrapped on a subspace of $X$, in particular
some holomorphic curve $C_i$, but can also wrap many times, so in
general has some non-negative multiplicity $a_i$. Thus we can conclude
that to describe a set of physical five-branes, $[W]$ must be the
homology class of what is called an effective curve. That is there
must be a representative curve in $[W]$ which is of the form  
\begin{equation}
   W = \sum_{i} a_{i}C_{i}
\label{curvedef}
\end{equation}
with irreducible holomorphic curves $C_i$ and integer multiplicities
$a_i$. 
\begin{itemize}
\item Effective condition: The requirement that $[W]$ is the class of
a set of physical five-branes, constrains $[W]$ to be effective. We
can show~\cite{future}, for this to be the case, that one must
guarantee  
\begin{equation}
  W_{B} \mbox{ is effective in $B$,} \quad
   a_{f}\geq0  \mbox{ integer } 
 \label{eq:x19}
\end{equation}
\end{itemize}
This requirement is a strong constraint on realistic non-perturbative
vacua.

Another obvious physical criterion for constructing realistic particle
physics models is that we should be able to find theories with a small 
number of families, preferably three. We will see that this is, in fact, 
easy to do via the bundle constructions on elliptically fibered 
Calabi--Yau three-folds that we are discussing. From the fact that all
the relevant fields are in the fundamental representation of fiber
group $SU(n)$, we have that the number of generations is
\begin{equation}
   N_{\text{gen}} = \text{index}\,(\Ds,V) 
      = \int_X \text{td}\,(X) \text{ch}\,(V)
      = \frac{1}{2}\int_X c_3(V)
\label{eq:x10}
\end{equation}
where $\text{td}\,(X)$ is the Todd class of $X$. For the case of
$SU(n)$ bundles on elliptically fibered Calabi--Yau three-folds, 
one can show, using equation~\eqref{c3} above and integrating
over the fiber, that the number of families becomes~\cite{Curio}
\begin{equation}
   N_{\text{gen}} = \l \eta ( \eta - n c_1(B) )
\label{eq:x11} 
\end{equation}
Restricting to three families and inserting equation~\eqref{eq:x17}
leads to the following condition on the vacua. 
\begin{itemize}
\item  Three family condition: The requirement that the theory have
three families imposes the  
constraint that 
\begin{equation}
   3 = \lambda \left( W_{B}^{2}- (24-n)W_{B}c_{1}(B)+
          12(12-n)c_{1}(B)^{2} \right)
\label{eq:x20}
\end{equation}
\end{itemize}
To these conditions, we reiterate the above constraint on $\lambda$ and $\eta$
rewritten, however, in terms of $W_{B}$. 
\begin{itemize}
\item  Bundle condition: The condition that $c_{1}({\cal{N}})$ be an
integer leads to the constraints on $W_{B}$ and $\lambda$ given by
\begin{equation}
\begin{aligned}
   {} & n \text{ is odd}, \quad \l = m+\frac{1}{2} \\
   {} & n \text{ is even}, \quad \l = m, 
      \quad W_{B} = c_1(B) \!\!\!\mod 2
\end{aligned}
\label{eq:x23}
\end{equation}
where $m$ is an integer.
\end{itemize}

It is important to note that all quantities and constraints have now been
reduced to properties of the base two-fold $B$. Specifically, if we know 
$c_1(B)$, $c_2(B)$, as well as a basis of effective classes in
$B$ in which to expand $W_{B}$, we will be able to exactly specify all
appropriate non-perturbative vacua. Hence, one proceeds as follows.
\begin{itemize}
\item  If we denote by $G=SU(n)$ the structure group of the gauge bundle
and by $H$ its commutant in $E_8$, then, for example
\begin{equation}
   G=SU(3) \Rightarrow  H=E_{6}, \quad
   G=SU(4) \Rightarrow  H=SO(10), \quad
   G=SU(5) \Rightarrow  H=SU(5)
\label{eq:x24}
\end{equation}
and $H$ corresponds to the low energy gauge group of the
theory. Choose the desired gauge group $H$ and, hence, the structure group
$G$. 
\item  Choose a base $B$ (an Enriques, a del Pezzo or a Hirzebruch
surface or its blow up) for the Chern classes $c_{1}(B)$ and
$c_{2}(B)$, as well as the effective classes.
\item  Specify $W_B$ and $\lambda$ subject to
effectiveness~\eqref{eq:x19}, three-family~\eqref{eq:x20} and
bundle~\eqref{eq:x23} constraints.  
\end{itemize}
We can use this prescription to produce numerous examples of three
family models with realistic gauge groups~\cite{future}. In this
letter, we will present one example which illustrates all the major
issues.

\subsection*{Example: $B=dP_{8}$}

We begin by choosing 
\begin{equation}
H=SU(5)
\label{eq:x25}
\end{equation} 
as the gauge group for our model. Then it follows from~\eqref{eq:x24}
that we must choose the structure group of the gauge bundle to be 
\begin{equation}
G=SU(5)
\label{eq:x26}
\end{equation}
and, hence, $n=5$. 

At this point, it is necessary to explicitly choose the base surface, which we
take to be
\begin{equation}
B=dP_{8}
\label{eq:x28}
\end{equation}
For the del Pezzo surface $dP_{8}$,  a basis
for $H_{2}(dP_{8}, \bf Z \rm)$ composed entirely of effective classes 
is given by $l$ and $E_{i}$ for $i=1,\dots,8$ with intersection numbers
\begin{equation}
l \cdot l=1  \qquad  l \cdot E_{i}=0  \qquad E_{i} \cdot E_{j}=-\delta_{ij}
\label{eq:x29}
\end{equation}
Furthermore, the first and second Chern classes of the $dP_{8}$ are given by
\begin{equation}
c_{1}(B)= 3l- \sum_{r=1}^{8} E_{i}, \qquad c_{2}(B)= 11
\label{eq:x31}
\end{equation}

We now must specify the component of the five-brane class in the
base $W_B$ and the coefficient $\l$, such that the
constraints~\eqref{eq:x19},~\eqref{eq:x20} and~\eqref{eq:x23} are
satisfied. Since $n$ is odd, equation~\eqref{eq:x23} tells
us that $\lambda=m+\frac{1}{2}$ for integer $m$. In this example we
will choose $m=1$ and $W_B$ so that 
\begin{align}
   W_{B} &= 2E_{1}+E_{2}+E_{3} \label{eq:x33} \\
   \lambda &= \frac{3}{2} \label{eq:x27}
\end{align}
Since $E_{1}$, $E_{2}$ and $E_{3}$ are effective, it follows that $W_{B}$ is
also effective, as it must be. Using the above intersection rules, one can
easily show that
\begin{equation}
   W_{B}^{2}=-6,  \qquad  W_{B}c_{1}(B)=4,  \qquad
   c_{1}(B)^{2}=1
\label{eq:x34}
\end{equation}
Using these results, as well as $n=5$ and $\lambda=\frac{3}{2}$, one
can show 
\begin{equation}
   a_{f} = c_{2}(B) 
            + c_1(B)^{2} \left(11+\frac{1}{24}(n^{3}-n) \right) 
            - \frac{3n}{2\lambda}\left(\lambda^{2}-\frac{1}{4} \right)
         = 17
\label{eq:x36}
\end{equation}
Since this is a positive integer, it follows from~\eqref{eq:x19} that
the full five-brane curve $[W]$ is effective, as it must be. Finally,
we see that 
\begin{equation}
   \lambda \left( W_{B}^{2} - (24-n)W_{B}c_{1}(B) 
      + 12(12-n)c_{1}(B)^{2} \right) = 3
\label{eq:x35}
\end{equation}
and, therefore, the three family condition~\eqref{eq:x20} is satisfied. 

This completes our construction of this explicit non-perturbative
vacuum. It represents a model of particle physics with three families
and gauge group $H=SU(5)$, along with  explicit five-branes wrapped on
a holomorphic curve specified by 
\begin{equation}
   [W]=2E_{1}+E_{2}+E_{3} + 17F
\label{eq:x37}
\end{equation}
We will now explore the properties of the five-branes and their moduli space
in detail.

The inclusion of five-branes not only generalizes the types of
bundles one can consider, but also introduces new degrees of
freedom into the theory, namely, the dynamical fields on the
five-branes themselves. From the point of view of a five-dimensional
effective theory on $M_4 \times S^1/Z_2$, since two of the five-brane
directions are compactified, it appears as a flat three-brane located
at some point $X^{11}$ on the orbifold. We have shown in previous
work~\cite{nse} that the low-energy four-dimensional theory
describing the five-brane dynamics will have $\cN=1$
supersymmetry. Furthermore, in general, the low-energy theory
of a single five-brane wrapped on a genus $g$ holomorphic curve $W$
has gauge group $U(1)^g$ with $g$ $U(1)$ vector multiplets and a
universal chiral multiplet with bosonic fields $(a,X^{11})$. The gauge
fields and axionic scalar field $a$ arise from the dimensional
reduction of the self-dual three-form $h$ on the five-brane
worldvolume. In additional, there are chiral multiplets describing the
moduli space of the curve $W$ in the Calabi--Yau three-fold.

This description of the low-energy five-brane is correct for a generic
five-brane. However, it is important to note that the gauge group can be
enhanced, either when two or more five-branes overlap or when a single
five-brane degenerates in such a way that two parts of the curve $W$
come close together in the Calabi--Yau manifold. In the following
analysis of the five-brane moduli space, we will see examples of both
types of enhancement. As an example, for $n$ overlapping five-branes,
each wrapped on the same genus $g$ curve $W$, the $n$ copies of
$U(1)^g$ are enhanced to $U(n)^g$. 

Within the context of the explicit three family, $H=SU(5)$ vacuum
discussed above, having so far only fixed the homology class $[W]$,
let us now analyze the moduli space of the curve $W$ in the
Calabi--Yau three-fold, as well as discuss the five-brane location
moduli $X^{11}$. We can construct the complete moduli space for this
theory. However, in this letter, we will present only a part of this
space which is sufficient to illustrate our main points. In general,
there are different branches of the moduli space corresponding to
different numbers of five-brane. The number of five-branes, and their
physical properties, will emerge from our analysis. In particular, we
will be able to exactly calculate the genus of the five-brane curves
and hence the resulting low-energy gauge group. 

The five-brane class $[W]$ in the above example is given by
\begin{equation}
[W]= 2E_1+E_2+E_3+17F
\label{eq:a1}
\end{equation}
We see that we have 17 copies of the fiber together with two copies of
$E_1$ and a single copy each of $E_2$ and $E_3$ in the base. 
Now, some of the $F$ part of the homology class can be shared among
the three distinct $E_{i}$ components. We are led to the
following generic decomposition 
\begin{equation}
\begin{split}
   [W] &= [W_1] + [W_2] + [W_3] + [W_4] \\
       &= \left(2E_1+aF\right) + \left(E_2+bF\right) 
          + \left(E_3+cF\right) + \left(17-a-b-c\right)F
\end{split}
\label{eq:a2}
\end{equation}
where $a$, $b$, and $c$ are integers. By this decomposition, we mean
that there are at least four separate five-branes (at least four
because each curve $W_i$ can further decompose into several
five-branes), except in the case where $17-a-b-c=0$ when there are at
least three. Physically this means that these five- branes lie at
arbitrary positions $X^{11}$ along the orbifold interval
$I=S^1/Z_2$. Note that for each of the four components to be
effective, we must restrict $a$, $b$, $c$ and $17-a-b-c$ to each be
non-negative, which we do henceforth. There are clearly many different
ways we can distribute $F$ among the three curves. Each distribution
corresponds to a different component of moduli space. We conclude that
the five-brane class in this specific example has a large moduli space
specified by the restricted set of integers $a$, $b$, and $c$. In
order to determine the exact number of five-branes and their physical
properties, we must examine the individual moduli spaces for each of
these four generic components. 

Let us first consider the curve
\begin{equation}
   [W_4] = \left(17-a-b-c\right)F \equiv f F
 \label{eq:a3}
\end{equation}
The complete moduli space for the curve $W_{4}$ (ignoring the axionic
scalar field $a$), for a fixed value of $f=17-a-b-c$, is given 
in the following table. 
\begin{equation}
\begin{array}{c|ccc}
   W_4 & \text{number of components} & \text{genus} & \text{moduli space} \\
       \hline
   fF  & 1 & 1 &dP_8\times I \\
   f_1 F + f_2 F & 2 & 1+1 & \left(dP_8\times I\right)^2 \\
       \vdots & \vdots & \vdots & \vdots \\
   F + \dots + F & f & f\times 1 & \left(dP_8\times I\right)^f
\end{array}
\end{equation}
where $f_1+\dots+f_n=f$ and we have not included the fact that if
there is a repetition of some of the $f_i$, we may have to mod out the
moduli space by a discrete group.
\begin{itemize}
\item Example: Consider the second line of the table. This
represents two independent five-branes lying at arbitrary positions 
$X^{11}$ along the orbifold interval $I$. Each five-brane is multiply
wrapped  on a holomorphic curve in $X$ with moduli space
$dP_{8}$. Since the genus of each holomorphic curve is unity, one might
expect a single $U(1)$ gauge field on each five-brane
worldvolume. However, the multiple wrapping means that the five-branes
are overlapping themselves. The result is that the gauge group on one
brane is enhanced to $U(f_1)$ whereas the other brane enhances to 
$U(f_2)$. This is an example of the second type of enhancement
mentioned above. Since the locations of the two branes are arbitrary,
there is a region of moduli space where they approach each other and
overlap. These overlapping branes further enhance the gauge group from
$U(f_1)\times U(f_2)$ to $U(f)$ where $f=f_{1}+f_{2}$. This is an
example of the first type of enhancement. As in the case of D-branes,
the whole moduli space of $W_4$ can be viewed as the moduli space of
vacua of a four-dimensional $U(f)$ theory. The different examples
listed above correspond to different Higgs branches of the theory.  
\end{itemize}

We now consider the $W_2$ and $W_3$ components. The construction of
the moduli space of $W_3$ proceeds in an identical fashion to that of
$W_2$. Hence, we will limit the discussion here to the class
$[W_2]$. This is given by  
\begin{equation}
   [W_2] = E_2 + bF
\label{eq:a4}
\end{equation}
As a concrete example, we will take $b=2$.
However, the analysis would be similar for any positive integer $b$. 
One can show that the curve $W_2$ must lie in a $dP_9$ surface
within the Calabi--Yau threefold. From our knowledge of del Pezzo
surfaces, we recall that a basis of curves on $dP_9$ are $l'$ and
$E'_i$ for $i=1,\dots,9$, where the prime distinguishes these curves
from their counterparts in the base $B=dP_{8}$ introduced above. A
subset of this moduli space is 
\begin{equation}
\begin{array}{c|ccc}
   W_2 & \text{number of components} & \text{genus} & \text{moduli space} \\
   \hline
   l'-E'_1-E'_2  & 
      1 & 0 & I \\
   2l'-E'_1-E'_2-E'_3-E'_4-E'_5 & 
      1 & 0 & I \\
   3l - \sum_{j=2}^9 E'_j &
      1 & 1 & \PP^1 \times I \\
   E'_2 + 2\left(3l-\sum_{j=1}^9E'_j\right) &
      3 & 1+0+1 & \PP^2 \times I^3
\end{array}
\end{equation}
\begin{itemize}
\item Example: Consider the first line of this table. This
represents one five-brane lying at some position $X^{11}$ along the
orbifold interval $I$. This five-brane is wrapped once around a
holomorphic curve in $X$ with a moduli space consisting of a single
isolated point with no freedom to deform within the Calabi--Yau
space. Since the genus of the holomorphic curve is $g=0$, it follows
that this brane has no gauge group on its worldvolume. We want to
emphasize this situation. By choosing the $W_{2}$ curve to lie at such
an isolated point of moduli space, one can dramatically reduce the
number of such moduli in the theory.  
\end{itemize}

Let us now briefly discuss the remaining component
\begin{equation}
   [W_{1}] = 2E_2 + aF
\label{eq:a6}
\end{equation}
For specificity, we consider the case $a=2$, although the analysis
would proceed in a similar way for any non-negative value of $a$. Part
of the moduli space for this case is given by 
\begin{equation}
\begin{array}{c|ccc}
   W_1 & \text{number of comp.} & \text{genus} & \text{moduli space} \\
   \hline
   2E'_2  & 
      2 & 0+0 & I \times I \\
   \hline
   l' - E'_1 &
      1 & 0 & \{\PP^1-8\text{ pts.}\} \times I \\
   (l' - E'_1 - E'_2) + E'_2 &
      2 & 0+0 & I \times I \\
   \hline
   3l' - \sum_{j=3}^9 E'_j &
      1 & 1 & \{\PP^2-{\cal{C}}\} \times I \\
   3l' - \sum_{j=3}^9 E'_j &
      1 & 0 & \{{\cal{C}}-21\text{ pts.}\} \times I \\
   (l'-E'_3-E'_4)+(2l'-E'_5-\dots-E'_9) &
      2 & 0+0 & I \times I 
\end{array}
\end{equation}
where ${\cal{C}}$ is the curve of zeros of the discriminant of the
cubic describing the curve. Note that within each divided set of rows,
a given curve in $dP_9$ decomposes into components. For example, the second
row describes a single five-brane which, at any one of eight points in
$\PP^1$, can split into the two independent five-branes described in
line three. Of particular interest is the last case, where the curve
is degenerating. This will be described in more detail
in~\cite{future}. 

\bigskip

We now turn to another physically important issue. That is, in the
non-perturbative vacua we are studying, how are the grand unified gauge
groups, such as $SU(5)$, $SO(10)$ and $E_{6}$, spontaneously broken to the
gauge group $SU(3)_{C} \times SU(2)_{L} \times U(1)_{Y}$ of the standard
model? From the decompositions of the $E_8$ gauginos given
in~\eqref{subgroup}, we see that the fermions in the adjoint
representation of the grand unified gauge group $H$ are singlets under
the structure group $G=SU(n)$. Thus, the low-energy zero-modes for these
fermions are all gauginos in the grand unified gauge supermultiplet
and, hence, there are no matter supermultiplets in the adjoint
representation. Therefore, there are no Higgs supermultiplets to break
the grand unified group $H$ to the standard model gauge group. To
achieve this symmetry breakdown, one must use Wilson lines, which
require the Calabi--Yau three-fold  $X$ to have a non-trivial
fundamental group. Does this occur for the non-perturbative vacua
discussed in this letter? 

It is most straightforward to discuss this issue within the context of
a specific non-perturbative vacuum, though we note examples for other
base manifolds can be constructed, in particular by relaxing the
condition that the fibration has a section~\cite{future}. Let us
consider an elliptically fibered Calabi-Yau three-fold with the base 
\begin{equation}
   B=K3/i
\label{eq:b1}
\end{equation}
where $i$ is the Enriques involution on $K3$ which takes the volume
form to minus itself. This base $B$ is an example of an Enriques
surface. The elliptically fibered Calabi-Yau three-fold with this base
has the structure 
\begin{equation}
 X= (K3 \times E)/(i, -1)
 \label{eq:b2}
\end{equation}
where $E$ is an elliptic curve and the action of $-1$ on $E$ is
inversion in the origin. Using the
methods of this letter, one can again construct bundles with fiber
group $G=SU(n)$ and realistic gauge groups $H$. Recall that $H$ is the
subgroup of $E_{8}$ that commutes with $SU(n)$. Now, one can show that
the first homotopy group $\pi_{1}(X)$ is the non-Abelian group
generated by the actions $(x,y) \rightarrow (x+1,y)$, $(x,y)
\rightarrow (x,y+1)$ and $(x,y) \rightarrow (-x,-y)$ in the real
plane. The group isomorphism 
\begin{equation}
 \pi_{1}/ (\ZZ \times \ZZ) \cong \ZZ_{2}
 \label{eq:b3}
\end{equation}
then implies that the structure group of the gauge bundle can be extended from
$G=SU(n)$ to $G=SU(n) \times \ZZ_2$. The low energy gauge group is now the
subgroup ${\cal{H}} \subset E_{8}$ which commutes with $G=SU(n) \times
\ZZ_{2}$. Clearly ${\cal{H}}$ is the subgroup of $H$ that commutes
with $\ZZ_2$ and we can think of the $\ZZ_2$ part of the structure group
as spontaneously breaking 
\begin{equation}
   H \longrightarrow {\cal{H}}
\label{eq:b4}
\end{equation}
For example, first choose the structure group $G=SU(5)$. It follows that $H$
is the grand unified group $H=SU(5)$. Now extend the bundle so that $G=SU(n) 
\times \ZZ_2$. The embedding of $\ZZ_2$ in $H=SU(5)$ can be
chosen so that 
\begin{equation}
 {\cal{G}}= \left( \begin{array}{cc}
              {\bf 1}_3 &  \\
              & - {\bf 1}_2 \\                    
              \end{array}
            \right )
\label{eq:b5}
\end{equation}
where ${\cal{G}}$ is the generating element of $\ZZ_2$. It follows
that the $\ZZ_2$ part of the fiber bundle spontaneously breaks
\begin{equation}
 SU(5) \longrightarrow SU(3)_{C} \times SU(2)_{L} \times U(1)_{Y}
 \label{eq:b6}
\end{equation}
and the standard model gauge group structure is achieved. 

\bigskip
{\bf Acknowledgments} 

\noindent
R.~D.~would like to thank Paul Aspinwall, Antonella Grassi, Mark Gross
and Tony Pantev. We would also like to thank Ed Witten for comments. 
R.~D.~is supported in part by an NSF grant DMS-9802456 as well as by a
University of Pennsylvania Research Foundation grant. A.~L. is
supported by the European Community under contract
No.~FMRXCT~960090. B.~A.~O.~is supported in part by DOE under contract
No.~DE-AC02-76-ER-03071 as well as by a University of Pennsylvania
Research Foundation grant . D.~W.~is supported in part by DOE under
contract No. DE-FG02-91ER40671. 

%%%%%%%%%%%%%%%%%%%%%%%%%%%%%%%%%%%%%%%%%%%%%%%%%%%%%%%%%%%%%%%%%%%%%%

%%%%%%%%%%%%%%%%%%%%%%%%%%%%%%%%%%%%%%%%%%%%%%%%%%%%%%%%%%%%%%%%%%%%%%%%%%%%

\end{document}